\documentclass[a4paper,11pt]{article}
\pdfoutput=1 

\usepackage{jinstpub} 

\usepackage{graphicx}
\usepackage{subfig}
\usepackage{svg}
\usepackage{amssymb}
\usepackage{amsmath}
\usepackage{lineno}
\usepackage{xcolor}
\usepackage{colortbl}
\usepackage{wrapfig}
\usepackage{textcomp}
\usepackage{lmodern,textcomp} 
\title{Design and evaluation of a versatile picosecond light pulser}

\author[a]{Martin Rongen\note{Corresponding author.}}
\author[a]{Merlin Schaufel}
\affiliation[a]{RWTH Aachen University, Institute for Particle Physics III B, 52074 Aachen, Germany}
\emailAdd{rongen@physik.rwth-aachen.de}

\abstract{
Many experimental setups, such as for the calibration of photosensors, require light sources with sub-nanosecond timing precision. We present the design and evaluation of a brightness controlled, picosecond light pulser based on an avalanche transistor pulse driving circuit. It can be used with a large variety of LEDs and VCSELs, enabling usage over a wide wavelength range, with light curves as short as $\sim$ 100\,ps standard deviation.
}


\begin{document}
\maketitle
\flushbottom

\section{Introduction}

In order to characterize photo-detectors with single photon timing resolutions in the picosecond to nanosecond range, such as photomultiplier tubes (PMTs) and silicon photomultiplier  (SiPM) \cite{SiPMTiming}, light sources with light curves faster than the studied detector are required.

Such sources are commercially available, but expensive. In contrast, simple circuits commonly used in the community, such as the Kapustinsky  pulser \cite{Kapustinsky1985}, are limited to light pulses of few nanosecond duration. 

The goal of this work is to describe the design and performance of a low-cost and easy to build light source that can supply sub-ns light pulses of variable intensity at an nearly arbitrary wavelength.

\section{The electric pulse driver}

\begin{figure}[!ht]
  \centering
  \includegraphics[width=0.5\columnwidth]{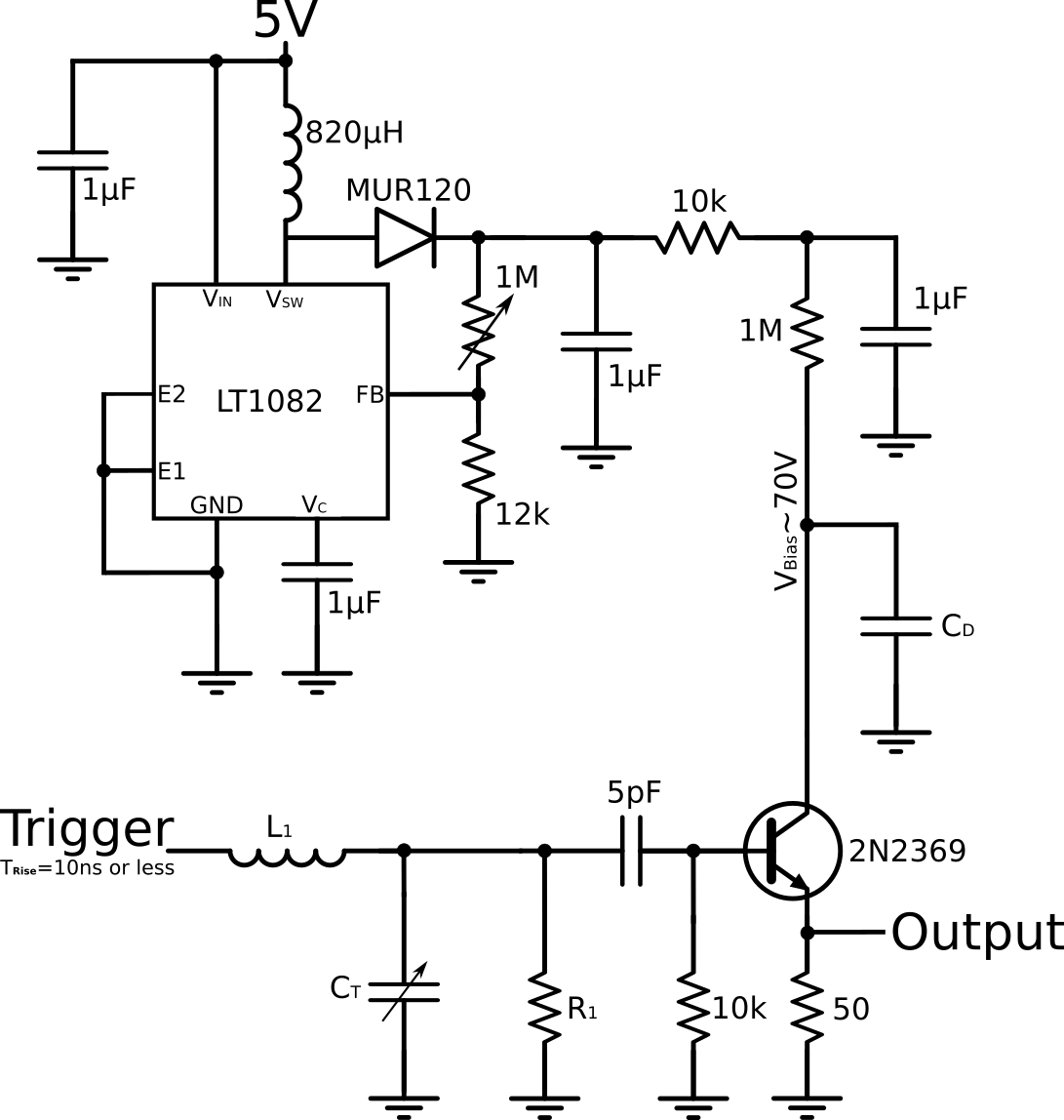}
  \caption{Triggered pulse generator as proposed in \cite{WilliamsPulser}}
  \label{FigWilliams}
\end{figure}

The central component of any fast light source is a circuit that can supply a sufficiently narrow electric drive pulse. For this purpose an avalanche transistor based circuit as described in \cite{WilliamsPulser} was chosen. The schematic is depicted in Figure \ref{FigWilliams}. The \textit{LT1082} switching regulator provides an adjustable bias voltage which is applied to a \textit{2N2369} avalanche transistor in reverse mode. The bias voltage (typically 70\,V) is chosen just below the breakdown point of the transistor such that no random breakdowns occur. An input pulse to the base can trigger an avalanche, discharging the discharge capacitance $C_D$ through the emitter. As the collector voltage drops the transistor self-resets and can be triggered again. The pulse height and width both scale with the discharge capacitance. For this study values of $\mbox{C}_{\mbox{D}}$ between 0\,pF (parasitics only) and 12\,pF were chosen. 

\begin{figure}[!ht]
  \centering
  \subfloat[Pulse driver prototype\label{FigPulserPCB}]{
  \includegraphics[width=0.4\columnwidth]{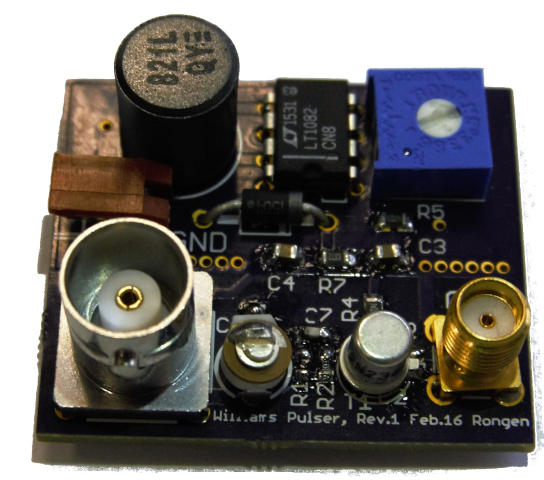}}
  \hfill
  \subfloat[Measured electric drive pulse\label{FigScope}]{
  \includegraphics[width=0.5\columnwidth]{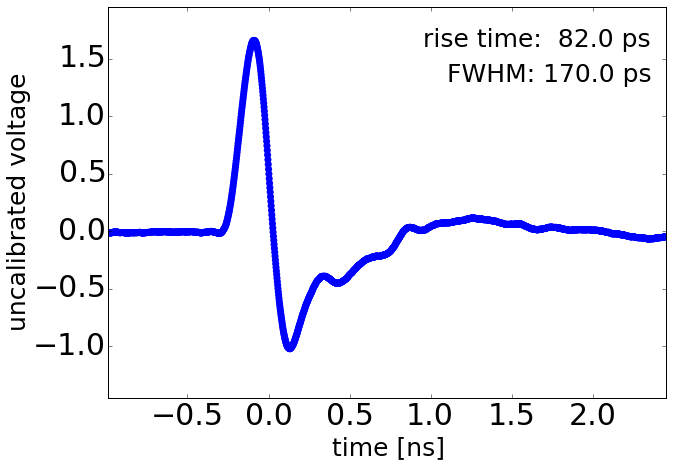}}
  \caption{PCB and voltage output of the electric pulse driver }
\end{figure}

The pulse shape of the drive circuit (as prototyped on a dedicated board, see Figure \ref{FigPulserPCB}) was evaluated using a \textit{Tektronix MSO71254C 100\,GS/s 12.5\,GHz Mixed Signal Oscilloscope} with a \textit{P7380A 8\,GHz} differential probe and single use solder contacts. As the probe only has a $\pm$2.5\,V dynamic range, a passive voltage divider was needed. Figure \ref{FigScope} shows a measured drive pulse, with only parasitic discharge capacitance in the circuit.

The pulse is bipolar, with a distinct, clean primary pulse of 170\,ps FWHM. The following undershoot does offer an advantage when applied to LEDs and solid state lasers, as it helps to remove charges from the depletion layer, thus reducing the turn-off time. This so called active sweep-out technique was for example suggested in \cite{SweepOut}.

While the overall bipolar shape has been observed reproducibly, the additional ringing in the tail changes significantly between repeated measurement with different voltage dividers and contact points. As such the ringing results at least partially from the measurement setup.

\section{Full light pulser module}

\begin{figure}[!ht]
  \centering
  \subfloat[Schematic layout of the light source \label{FigFullSchematic}]{
  \includegraphics[width=0.65\columnwidth]{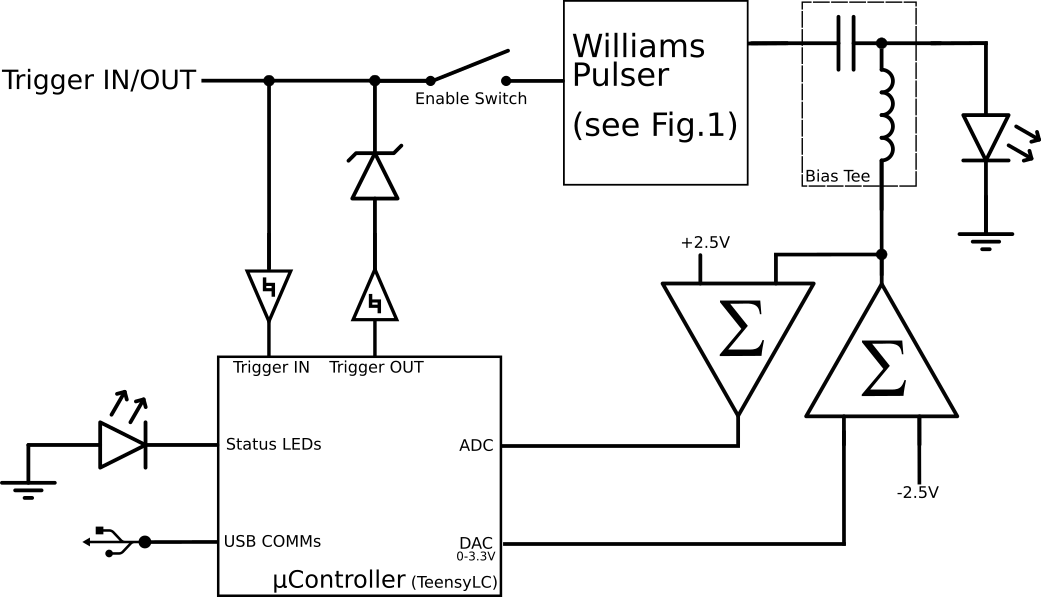}}
  \hfill
  \subfloat[Photo of a fully assembled module\label{FigModule}]{
  \includegraphics[width=0.33\columnwidth]{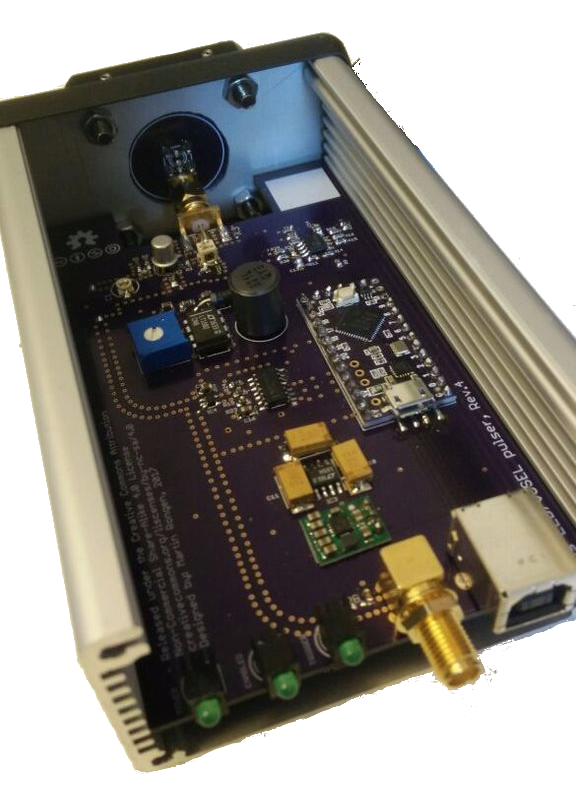}}
  \caption{Hardware overview of a fully integrated picosecond light pulser module.}
\end{figure}

In order to provide a simple to use light source the electric pulse driver as described in the previous section is incorporated into stand-alone light pulser module (see figure \ref{FigModule}). It provides an additional micro-controller (\textit{Teensy LC}) for trigger control and a \textit{Mini-Circuits TCBT-14+ 10\,MHz-10\,GHz} bias tee to shift the DC level of the drive pulse. The light source is attached via a SMA breakout board, so that different sources can be easily used with the same module. Tested light sources are an \textit{OPV332} 850\,nm vertical cavity surface emitting laser (VCSEL) from TT Electronics \cite{VCSEL}
 (expected rise/fall time $\sim 100\,ps$) and several 3\,mm through-hole LEDs purchased from \textit{Roithner Lasertechnik}. 

A trigger pulse can either be provided externally, with the micro-controller monitoring the frequency, or can be generated by the micro-controller directly. In both cases the light output can be inhibited by an analog switch to the pulse driver trigger input.

An offset voltage between -2.5\,V and 0.8\,V is generated by a summing amplifier which is connected to a -2.5\,V reference and a DAC output. This offset voltage is applied to the drive pulse via the bias-tee.
This allows for an effective intensity control, without changing the shape of the drive pulse. As the integral of the drive pulse over the forward voltage, as well sweep-out efficiency changes with biasing, the width of the light curve also changes. Negative bias voltages result in a narrower but dimmer light curve.

To achieve the highest possible intensity, no additional current limiting resistor is placed in series with the LED/VCSEL. This is safe as the total energy of the pulse is limited by the discharge capacitor $C_D$. In fact, depending on  the biasing, the voltage of the LEDs is barely over the forward voltage and no aging has been observed for either LEDs or VCSELs over many days of continuous operation at high repetition frequencies.

The total material cost for the module including an extruded aluminum enclosure and a one inch optics adapter is $\sim$ 150\texteuro. 

\section{Timing characterization}

The light source is being characterized in terms of the profile of the light curve, the trigger delay as well as the integral photon output.

\subsection{APD setup}

The measurement of the light curve is obtained by using a 50\, \textmu m avalanche photodiode based \textit{IDQ ID-100} APD detector \cite{ID100BuH}. The device outputs a 10\,ns, 2\,V rectangular pulse when one or more incident photons are detected. The single photon timing resolution is guaranteed to be smaller then 60\,ps FWHM and is usually below 40\,ps \cite{ID100Paper}. 

In this setup, the light curve is measured as the distribution of timing delays between the internal trigger pulse of the light source and the arrival time of single photons at the APD. For this purpose a \textit{TDC7200 evaluation board} \cite{TDC} is being used as Time to Digital Converter. As the APD can not distinguish the number of registered photons, the occupancy, that is the increase in APD detection rate relative to the trigger rate, needs to be small ($\sim$10\,\%) to guarantee mostly single photons. In order to measure the occupancy the APD rate is monitored with a custom frequency counter, consisting of a \textit{SN74LVC} monostable multi-vibrator and a micro-controller.

\begin{figure}[!ht]
  \centering
  \subfloat[Timing resolution when measuring a precise delay\label{FigTDCResolution}]{
  \includegraphics[width=0.45\columnwidth]{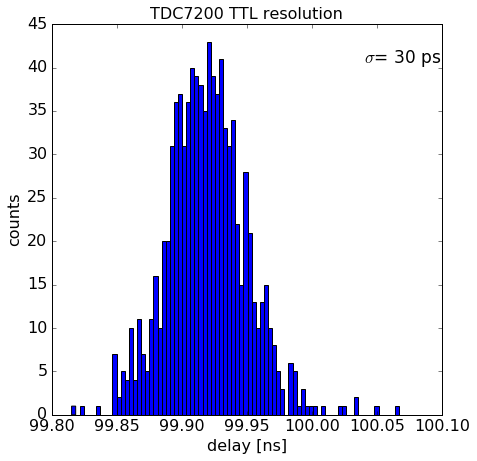}}
  \hfill
  \subfloat[TDC7200 binning behavior for broad time spreads\label{FigTDCBinning}]{
   \includegraphics[width=0.45\columnwidth]{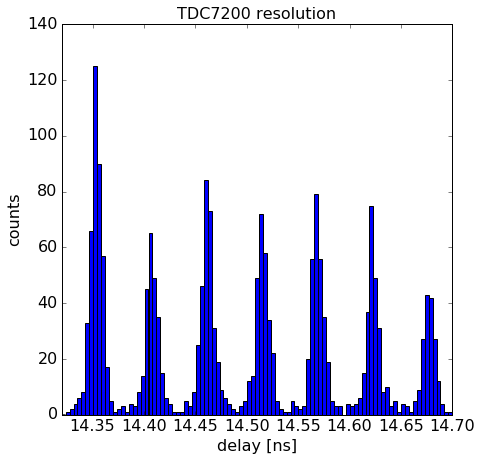}}
  \caption{Performance of the TDC7200 evaluation board}
\end{figure}

The TDC7200 measures delays, by counting the number of clock cycles of a multiple-GHz ring oscillator, between a rising/falling edge of a start and a stop signal. As the frequency of ring oscillators is hard to determine and can drift over time, it is calibrated against an external, precision 8\,MHz clock after each trigger. The TDC7200 datasheet \cite{TDC} claims a resolution of 55\,ps (one cycle of the ring oscillator) and a standard deviation of 35\,ps (calibration uncertainty). 

The timing accuracy of the TDC7200 has been verified by measuring two TTL pulses at 100\,ns delay as provided by a \textit{Quantum Composer 9518+} pulse generator. The standard deviation of the timing delay as seen in Figure \ref{FigTDCResolution} agrees with the specification. When measuring the delay of a less well determined signal (see Figure \ref{FigTDCBinning}), the fact that the resolution is lower than the standard deviation becomes apparent as "fingers" in the histogram. In the following the binning of all delay histograms has been chosen to match the TDC resolution.

\subsection{IR-VCSEL timing performance}

\begin{figure}[!ht]
  \centering
\subfloat[compared to the APD IRF\label{FigVCSELSiPM}]{
  \includegraphics[width=0.45\columnwidth]{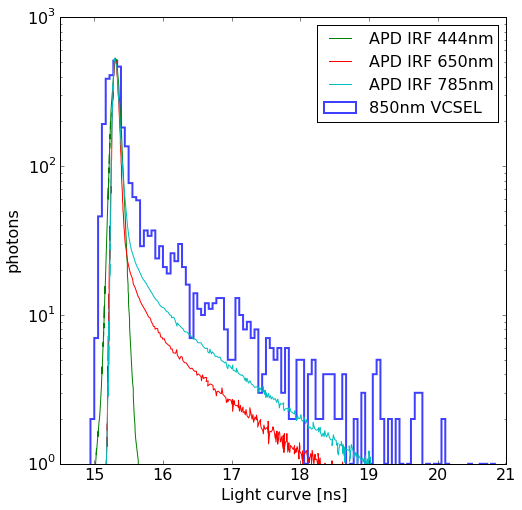}}
  \hfill  
 \subfloat[after cleaning away the APD IR diffusion tail \label{VCSELcleaned}]{
     \includegraphics[width=0.45\columnwidth]{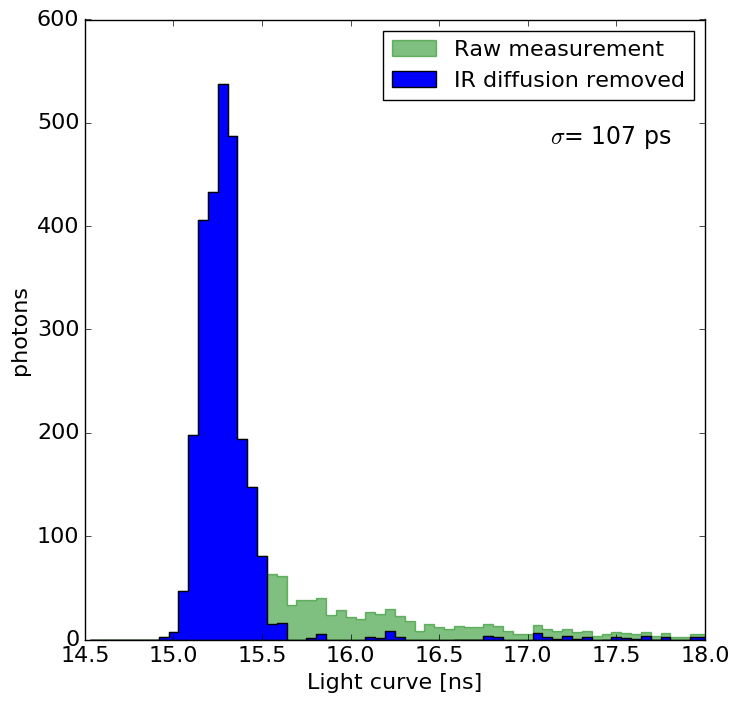}} 
  \caption{VCSEL light curve at -2.5\,V biasing with no discharge capacitance}
\end{figure}

Out of the tested light sources the VCSEL achieves the best timing performance. Figure \ref{FigVCSELSiPM} shows the measured light curve compared to the expected APD instrument response function (IRF) at different wavelengths. In addition to a primary, short duration peak a long duration tail is observed. This is most likely due to an APD detector effect, as APDs suffer from significant infrared diffusion tails, limiting the ability to quantify potential, low brightness, long duration turn-off tails of the light curves at long wavelengths \cite{ID100BuH}. 

As no exact APD IRF is available at the VCSEL wavelength of 850\,nm the IR diffusion tail is fitted to the data assuming a simple exponential behavior. Figure \ref{VCSELcleaned} shows the cleaned light curve after subtracting the IR diffusion component. The light pulser output shows a clean Gaussian profile with a standard deviation of $\sim 100$\,ps.

\subsection{LEDs timing performance}

As the wavelength availability of VCSELs is very limited, a selection of 26 LEDs ranging from 365\,nm to 810\,nm was tested. All LEDs are 3\,mm through-hole variants ordered from \textit{Roithner Lasertechnik}. These LEDs are not designed for pulsed applications and no rise-times are stated in the datasheets. 

As the emitted intensity has been observed to be significantly lower compared to VCSEL measurements, the APD setup for timing measurements had to be adapted. Instead of starting the TDC on the rising edge of the trigger pulse, it is started on an APD signal and stopped on the falling edge of the trigger pulse. The occupancy is still checked with the frequency counter but is always below 1\,\%. As the width of the trigger pulse is not perfectly controlled ($\sigma\sim 60$\,ps) the single photon timing resolution of this configuration is slightly worse at $\sim 80$ \,ps.

\begin{figure}
\begin{minipage}[c]{0.55\linewidth}
\centering
\includegraphics[width=0.9\linewidth]{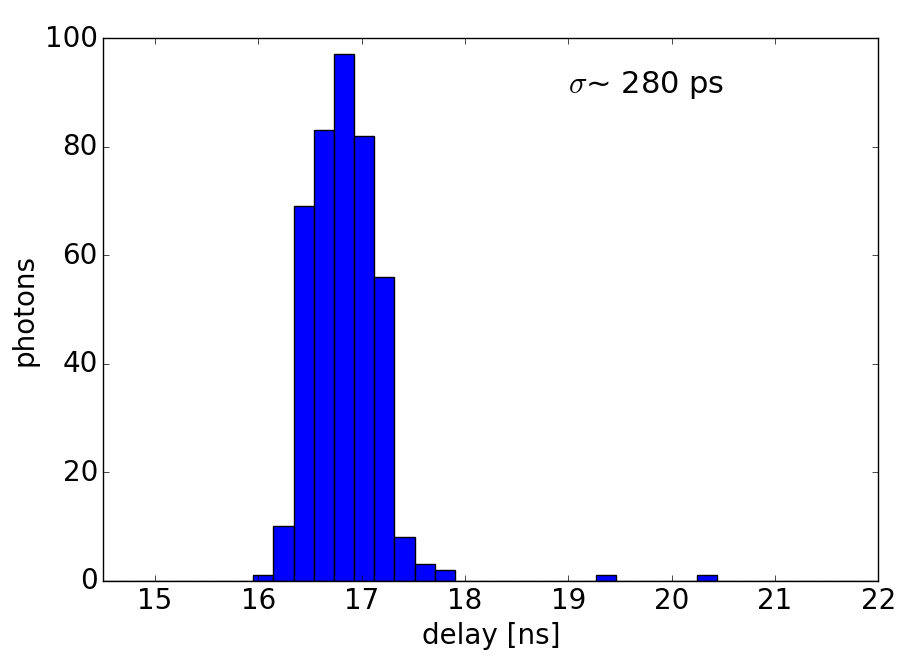}
  \caption{385nm LED light curve at -2.0\,V biasing and 10\,pF discharge capacitance}
  \label{APD_385}
\end{minipage}
\hfill
\begin{minipage}[c]{0.4\linewidth}
  \centering
  \includegraphics[width=0.9\linewidth]{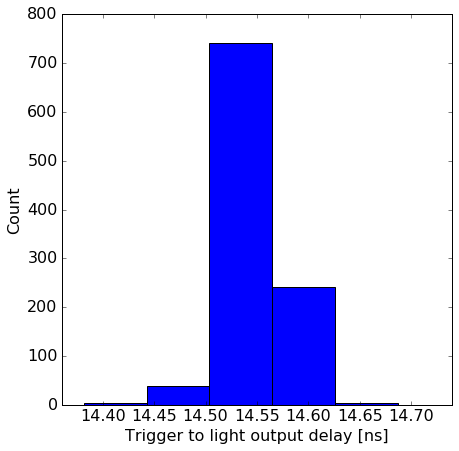}
  \caption{Trigger stability}
  \label{FigTrigger}
\end{minipage}%
\end{figure}

Only 20 LEDs emitted a detectable amount of light. Of these eight (370\,nm, 375\,nm , 385\,nm, 590\,nm, 605\,nm, 680\,nm, 700\,nm and 770\,nm) exhibit light curves shorter then 1\,ns standard deviation. In general high wavelength LEDs are easier to pulse due to the smaller band gaps, yet the best light curve was obtained with a 385\, nm LED as shown in Figure \ref{APD_385}. For the non-IR LEDs no long-duration tail is observed, confirming that the effect seen for the VCSEL is indeed a detector artifact caused by the APD.

\section{Trigger stability characterization}

The trigger stability is defined as the standard deviation of the timing delay between trigger pulses and the beginning of the light curve. As such it includes both: the jitter between the trigger and the drive pulse as well as possible turn-on variations of the LED/VCSEL. It is measured by illuminating the APD at the maximum possible brightness. As several photons reach the APD nearly simultaneously, this measurement is not limited by the APD single photon timing resolution. No dependence of the trigger stability on the used light source was found. A histogram of the registered timing delays is shown in figure \ref{FigTrigger}. The spread is barely detectable at $\sim 40$\,ps and matches the spread seen when measuring the jitter of the electric drive pulse.

\section{Intensity characterization}

In order to quantify the pulse intensity (photons per pulse) a \textit{Hamamatsu S2281} photodiode has been used in conjunction with a custom preamplifier as described in \cite{Tosi:2015ica}. While the intensity of a single pulse is not sufficient to be distinguished from the noise of the photodiode system, the average pulse intensity can be measured as the slope of the dark noise corrected intensity versus the light source repetition frequency.

\begin{figure}[!ht]
  \centering
  \includegraphics[width=0.6\columnwidth]{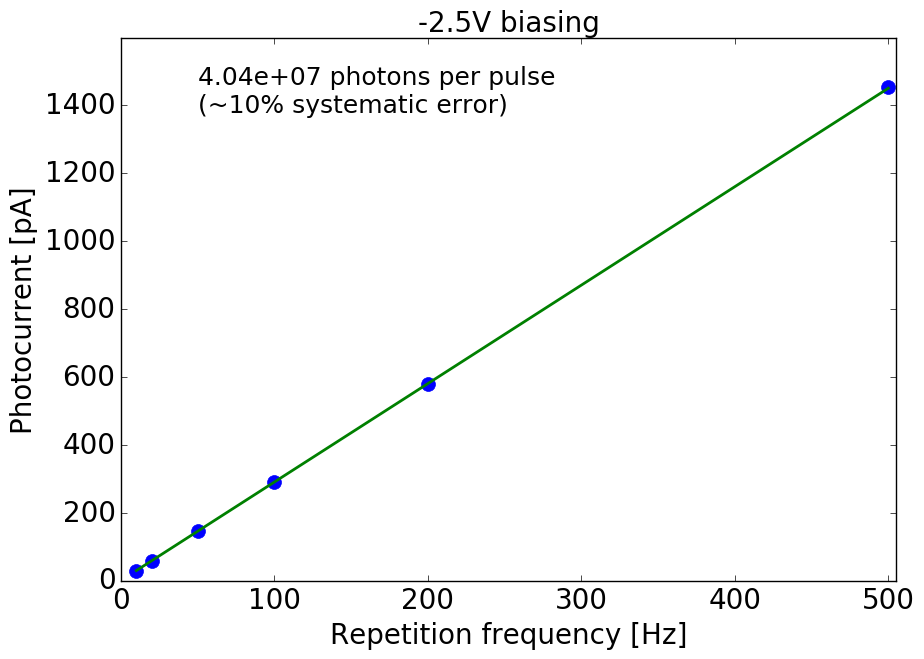}
  \caption{Example measurement of the VCSEL 0\,pF per-pulse light output at -2.5\,V biasing}\label{FigVCSELS2281}
\end{figure}

Figure \ref{FigVCSELS2281} shows the measured photo-current as a function of repetition frequency for a VCSEL with 0\,pF discharge capacitance at -2.5\,V biasing. The slope and the quantum efficiency from \cite{S2281} yield $\sim 4 \cdot 10^7$ photons per pulse. An additional 10\,\% systematic error on the quantum efficiency has to be considered. 

Increasing the bias voltage to 0.7\,V results in a threefold increase of the photon number at the expense of a slightly longer light curve.

For the LEDs, pulse intensities between $10^4$ photons per pulse (no discharge capacitance, low biasing) and $10^8$ photons per pulse (10\,pF  and positive biasing) are seen. For the earlier mentioned 385\,nm LED at 10\,pF the photon number goes from $\sim 5\cdot 10^4$ photons per pulse at -2.0\,V biasing to $\sim 10^7$ photons per pulse at 0.5\,V biasing. This leads to an increase in the light curve standard deviation from $\sim$280\,ps to $\sim$800\,ps

\section{Summary}
Using a well established avalanche transistor based pulse driving circuit, combined with biasing and trigger control a versatile, low-cost and easy to use light source has been build and characterized. Using LEDs nearly arbitrary wavelengths can be realized with sub-ns timing precision. In conjunction with a 850\,nm VCEL light curves as short as 100\,ps  have been achieved.

\acknowledgments
This work was supported in part by BMBF, Verbundforschung and was carried out in the context of the IceCube Neutrino Observatory.
We thank the ZEA-2, Forschungszentrum J\"ulich for the possibility to use their 100\,GS/s oscilloscope. We would also like to thank Christopher Wiebusch and Heinz Rongen for their support of this project.

\bibliography{sample}

\end{document}